\documentclass[12pt]{article}
\usepackage{amssymb,amsmath}
\begin{document}

\baselineskip=.3in
\renewcommand{\baselinestretch}{1.5}

\title{\Large \bf Some exact solutions of 
the semilocal Popov equations
\vspace{10mm}}
\author{Chanju Kim\thanks{cjkim@ewha.ac.kr}\\[5mm]
{\normalsize\it Department of Physics and Institute for the Early Universe,}\\
{\normalsize\it Ewha Womans University, Seoul 120-750, Korea}
}
\date{}
\maketitle

\vspace{10mm}

\begin{abstract}
We study the semilocal version of Popov's vortex equations on $S^2$.
Though they are not integrable, we construct two families of exact solutions
which are expressed in terms of rational functions on $S^2$.
One family is a trivial embedding of Liouville-type solutions of the Popov
equations obtained by Manton, where the vortex number is an even integer.
The other family of solutions are constructed through a field redefinition
which relate the semilocal Popov equation to the original Popov equation 
but with the ratio of radii $\sqrt{3/2}$, which is not integrable. These
solutions have vortex number $N=3n-2$ where $n$ is a positive integer,
and hence $N=1$ solutions belong to this family.
In particular, we show that the $N=1$ solution with reflection symmetry 
is the well-known $CP^1$ lump configuration with unit size where the scalars 
lie on $S^3$ with radius $\sqrt{3/2}$. It generates the
uniform magnetic field of a Dirac monopole with unit magnetic charge on $S^2$.
\end{abstract}


\newpage

Recently, Popov \cite{Popov:2012av} obtained a set of vortex-type equations on 
a 2-sphere by dimensional reduction of SU(1,1) Yang-Mills instanton 
equations on the four-manifold $S^2 \times H^2$ where $H^2$ is a hyperbolic 
plane.  It was shown that they
are integrable when the scalar curvature of the manifold vanishes.  
Subsequently, Manton \cite{Manton:2012fv} constructed explicit solutions with
even vortex numbers from rational functions on the sphere. They
have a geometric interpretation in terms of conformal rescalings 
of the 2-sphere metric.

The Popov equations involve a complex scalar field and a U(1)
gauge potential. Except a flipped sign, they are the same as the well-known 
Bogomolny equations \cite{Bogomolny:1975de} for abelian Higgs vortices 
\cite{Abrikosov:1956sx,Nielsen:1973cs} on $S^2$.
In this paper, we would like to consider the semilocal 
\cite{Vachaspati:1991dz,Hindmarsh:1991jq}
version of the Popov equations, which consist of two scalar fields instead
of one. The equations have an additional global SU(2) symmetry with respect to
the rotation of the scalars as well as the local U(1) symmetry. We will
show that they appear in 2+1 dimensional Chern-Simons systems with
nonrelativistic matter on $S^2$. Such systems on the plane have been
extensively studied to understand quantum Hall effect and other related
phenomena \cite{Wilczek,Jackiw:1990tz,Ezawa:1991sh}.
Then we construct two families of exact solutions of the semilocal
Popov equations. One family of solutions is trivially
obtained by a simple ansatz that the two scalars are proportional to each
other, with which the equations reduce to the original Popov equations.
For the other family of solutions, we will relate the equations to the semilocal
version of the Liouville equations considered in \cite{Kim:1992uw,Kim:1993mh}. 
In addition to Liouville solutions, they admit another family of exact
solutions \cite{Kim:1993mh} which involves an arbitrary rational function 
on $S^2$.
We will construct solutions of the semilocal Popov equations from them.

It turns out that semilocal Popov equations have another 
connection to the original Popov equations with a single scalar. As mentioned
above, it is integrable only when the scalar curvature of the underlying
four manifold $S^2 \times H^2$ vanishes \cite{Popov:2012av}, which happens for 
equal radii $R_1=R_2$, where $R_1$, $R_2$ are the radii of $S^2$ and $H^2$,
respectively. Here we will show that the semilocal Popov equations with 
equal radii can be transformed to the Popov equations with different radii
$R_1/R_2 = \sqrt{3/2}$. The aformentioned solutions of the semilocal
equation correspond to the constant solution of the latter.

The Liouville solutions have only even vortex numbers \cite{Manton:2012fv}. 
However the vortex number of the other family of solutions is $N=3n-2$, 
where $n$ is a positive integer, so that odd vortex numbers are possible. 
In particular, the solutions
with unit vorticity $N=1$ belong to this family. We will show that the $N=1$
solution with reflection symmetry in the equator of $S^2$ is
precisely given by the $CP^1$ lump configuration with unit size. The
$S^3$ where the scalar fields lie has radius $\sqrt{3/2}$ which is the ratio
$R_1/R_2$ above. The magnetic field is that of a Dirac monopole with 
unit magnetic charge on $S^2$.

Let us begin with writing the Popov equations on $S^2$.
For convenience, the radius of $S^2$ is fixed to be $\sqrt2$. The metric
of $S^2$ is given by $ds^2 = \Omega dzd\bar z$ with
\begin{equation} \label{Omega}
\Omega = \frac{8}{(1+|z|^2)^2}.
\end{equation} 
The Popov equations are\footnote{We follow the notation 
of \cite{Manton:2012fv}.}
\begin{align}
D_{\bar z} \phi &\equiv \partial_{\bar z} \phi - i a_{\bar z} \phi = 0 \,, 
\label{Popov1} \\
F_{z \bar z} &=- \frac{2i}{(1 + |z|^2)^2} ( C^2 - |\phi|^2 ) \,. \label{Popov2}
\end{align}
where $C=R_1/R_2 = \sqrt2 / R_2$ is the ratio of radii as described above.
$\phi$ is a complex scalar field, $a$ is a U(1) gauge 
potential and $F_{z\bar z} = \partial_z a_{\bar z} - \partial_{\bar z} a_z$ 
is the field strength which is imaginary. 
If the right hand side of the second equation has opposite sign, these would
be the same as the Bogomolny equations for abelian Higgs vortices on $S^2$.
As mentioned above, the equations are integrable only for $C=1$ 
\cite{Popov:2012av}. From \eqref{Popov1} the gauge potential $a_{\bar z}$ may
be expressed as
\begin{equation} \label{az}
a_{\bar z} = -i \partial_{\bar z} \ln \phi,
\end{equation}
away from zeros of $\phi$.
Since 
\begin{equation}
	F_{z \bar z} = -i \partial_z \partial_{\bar z} \ln |\phi|^2,
\end{equation}
we can eliminate the gauge potential and are left with a single equation
\begin{equation} \label{Popov3}
\partial_z \partial_{\bar z} \ln |\phi|^2
        = \frac2{(1+|z|^2)^2} (C^2 - |\phi|^2),
\end{equation}
which is valid away from zeros of $\phi$.

Equations \eqref{Popov1} and \eqref{Popov2} may be obtained from an
energy function \cite{Popov:2012av,Manton:2012fv} which comes from
a dimensional reduction of the Yang-Mills action,
\begin{align} \label{energy}
E &= \frac12 \int_{S^2}  \left[ \frac4\Omega |F_{z \bar z}|^2
    -2(|D_z\phi|^2 + |D_{\bar z}\phi|^2 )
    + \frac\Omega4 (C^2 - |\phi|^2)^2 \right] 
      \frac{i}2 dz \wedge d\bar z \notag \\
 &= \frac12 \int_{S^2}  \left\{ -\frac4\Omega
     \left[ F_{z \bar z} + i\frac{\Omega}4 (C^2 - |\phi|^2) \right]^2
     -4 |D_{\bar z}\phi|^2 \right\} \frac{i}2 dz \wedge d\bar z -\pi C^2 N \,,
\end{align}
where $N$ is the first Chern number
\begin{equation}
N = \frac1{2\pi} \int_{S^2} F_{z\bar z} dz \wedge d\bar z,
\end{equation}
which is an integer and is the same as the vortex number which counts
the number of isolated zeros of $\phi$. 
Therefore, for fields satisfying the Popov equations, 
the energy is stationary and has value $-\pi C^2 N$.
It is however not minimal because of the negative sign in the second term
of \eqref{energy}.

The Popov equation \eqref{Popov3} can also arise in a completely different 
physics system.
Let us consider a $2+1$ dimensional Chern-Simons gauge theory with 
a nonrelativistic matter field on $S^2$ of which the action is
\begin{equation} \label{CS1}
S = \int dt \int_{S^2} \left[ 
  \frac\kappa2 \epsilon^{\mu\nu\lambda} a_\mu \partial_\nu a_\lambda
 + \Omega (i \phi^* D_t \phi - V)
 - (|\tilde D_z\phi|^2 + |\tilde D_{\bar z}\phi|^2 )
  \right] \frac{i}2 dz \wedge d\bar z,
\end{equation} 
where $\kappa$ is the Chern-Simons coefficient and
\begin{align}
D_t \phi &= (\partial_t - i a_t ) \phi \, \notag \\
\tilde D_z \phi &= (\partial_z - i a_z - iA_z^{ex}) \phi \,.
\end{align}
Note that we applied an external $U(1)$ gauge potential $A^{ex}$ given by
\begin{equation}
A_{\bar z}^{ex} = \frac{i}2 \frac{gz}{1+|z|^2}, 
\end{equation}
which generates uniform magnetic field with magnetic charge $g$ on $S^2$.
The potential $V$ has the form
\begin{equation} \label{potential}
V = -\frac{g}8 |\phi|^2 + \frac1{2\kappa} |\phi|^4.
\end{equation}
This action has been extensively studied on the plane 
in the context of anyon physics
to understand quantum Hall effect and other related phenomena
\cite{Jackiw:1990tz,Ezawa:1991sh}.

Variation of $a_t$ gives the Gauss constraint
\begin{equation} \label{gauss}
F_{z \bar z} = -i \frac\Omega{2\kappa} |\phi|^2.
\end{equation}
The energy function is
\begin{equation}
E = \int_{S^2} ( |\tilde D_z\phi|^2 + |\tilde D_{\bar z}\phi|^2  + \Omega V )
           \frac{i}2 dz \wedge d\bar z,
\end{equation}
which has no explicit contribution from the Chern-Simons term.
It can be rewritten by the usual Bogomolny rearrangement
\begin{align} \label{added}
|\tilde D_z \phi |^2 &= |\tilde D_{\bar z} \phi |^2
                -i( F_{z \bar z} + F_{z \bar z}^{ex} ) |\phi|^2 \notag \\
   &= |\tilde D_{\bar z} \phi |^2
   - \frac{\Omega}{2\kappa} |\phi|^4 + \frac{g}8 \Omega |\phi|^2,
\end{align}
up to a total derivative term, where in the second line we have used \eqref{gauss} and 
$F_{z \bar z}^{ex} = \frac{ig}8 \Omega$. The last two terms in \eqref{added}
are cancelled by the potential \eqref{potential} and the energy becomes
\begin{equation}
E = 2 \int_{S^2} |\tilde D_{\bar z}\phi|^2\, \frac{i}2 dz \wedge d\bar z,
\end{equation}
which is positive definite. Therefore the energy vanishes if
\begin{equation}
\tilde D_{\bar z}\phi = 0.
\end{equation}
Combining this equation with the Gauss constraint \eqref{gauss}, we get
\begin{equation}
\partial_z \partial_{\bar z} \ln |\phi|^2
  = - \frac{\Omega}{2\kappa} \left( \frac{\kappa g}4 - |\phi|^2 \right),
\end{equation}
away from zeros of $\phi$. With $\kappa = -2$ and $g = -2C^2$ 
this becomes the Popov equation \eqref{Popov3}.

Now we introduce the semilocal Popov equations which involve
two scalar fields $\phi_i$ ($i=1,2$). We will only consider
the case of equal radii, i.e., $C=1$. The semilocal Popov equations read
\begin{align}
D_{\bar z} \phi_i &\equiv \partial_{\bar z} \phi_i - i a_{\bar z} \phi_i = 0 \,, 
\qquad (i=1, 2) \label{sPopov1} \\
F_{z \bar z} &=- \frac{2i}{(1 + |z|^2)^2}(1 - |\phi_1|^2 - |\phi_2|^2) \,,
\label{sPopov2}
\end{align}
which have an obvious global SU(2) symmetry. 
These equations can again be obtained from the energy function generalizing
\eqref{energy} by introducing two scalars
\begin{align}
E &= \frac12 \int_{S^2}  \left\{ \frac{(1 + |z|^2)^2}2 |F_{z \bar z}|^2
    -2\sum_{i=1}^2(|D_z\phi_i|^2 + |D_{\bar z}\phi_i|^2 )
+\frac{2}{(1 + |z|^2)^2}(1 - |\phi_1|^2 - |\phi_2|^2)^2 \right\} 
      \frac{i}2 dz \wedge d\bar z \,.
\end{align}
Another way is to consider the action
\begin{equation} \label{CS2}
S = \int dt \int_{S^2} \left\{ 
  \frac\kappa2 \epsilon^{\mu\nu\lambda} a_\mu \partial_\nu a_\lambda
 + \Omega (i \bar\phi_1 D_t \phi_1 + i \bar\phi_2 D_t \phi_2 - V)
 - \sum_{i=1}^2 (|\tilde D_z\phi_i|^2 + |\tilde D_{\bar z}\phi_i|^2 )
  \right\} \frac{i}2 dz \wedge d\bar z,
\end{equation} 
with
\begin{equation}
V = -\frac{g}8 (|\phi_1|^2 + |\phi_2|^2)
     + \frac1{2\kappa} (|\phi_1|^2 + |\phi_2|^2)^2.
\end{equation}
This class of theories were considered on the plane to study double-layer
electron systems \cite{Ezawa:1991sh}.
Now following the same procedure as above, it is straightforward to get 
the semilocal Popov equations.

A Bradlow-type constraint \cite{Bradlow:1990ir} on $N$ can be obtained by
integrating \eqref{sPopov2} over the sphere of radius $\sqrt2$,
\begin{equation} \label{Bradloweq}
\int_{S^2}  \Omega (|\phi_1|^2 + |\phi_2|^2) 
\frac{i}2 dz \wedge d{\bar z} = 8\pi + 4\pi N \,,
\end{equation}
which implies $N \ge -2$.

Let us now consider the first equation \eqref{sPopov1} which represents
the semilocal nature of the equations. As before it can be written as
\begin{equation} \label{az2}
a_{\bar z} = -i \partial_{\bar z} \ln \phi_i
\end{equation}
for both $\phi_1$ and $\phi_2$. Then
\begin{equation}
\partial_{\bar z} \ln \left( \frac{\phi_2}{\phi_1} \right) = 0,
\end{equation}
so that the ratio 
\begin{equation} \label{wz}
w(z) = \frac{\phi_2}{\phi_1}
\end{equation}
is locally holomorphic. Moreover, \eqref{sPopov1} implies that $\phi_i$'s have
zeros at discrete points \cite{Taubes:1979tm}. Therefore $w(z)$ should be
a rational function of $z$. We can again eliminate $a_{\bar z}$ 
and \eqref{sPopov2} becomes
\begin{equation} \label{sPopov3}
\partial_z \partial_{\bar z} \ln |\phi_1|^2
    = \partial_z \partial_{\bar z} \ln |\phi_2|^2
    = \frac2{(1+|z|^2)^2} (1 - |\phi_1|^2 - |\phi_2|^2 ),
\end{equation}
away from zeros of $\phi_i$. \eqref{sPopov3} is the same as the semilocal
abelian Higgs vortex equations \cite{Vachaspati:1991dz,Hindmarsh:1991jq}
on $S^2$ except the flipped sign, as it should be. It is illuminating to introduce 
\begin{equation} \label{uphi}
e^{u_i} = \frac{|\phi_i|^2}{(1+|z|^2)^2},
\end{equation}
and rewrite \eqref{sPopov3} as a Toda-type equation,
\begin{equation} \label{toda}
\partial_z \partial_{\bar z} u_i =  -K_{ij} e^{u_j},
\qquad K = \begin{pmatrix} 2 & 2 \\ 2 & 2 \end{pmatrix}.
\end{equation}
It is known to be integrable if the matrix $K$ corresponds to the Cartan matrix
of a Lie algebra \cite{Leznov:1979td,Konstant}, 
which is not the case for \eqref{toda}. It may be considered
as the semilocal version of the Liouville equation. Although it
is not integrable, some exact solutions have been obtained in the context 
of nonrelativistic self-dual Chern-Simons matter systems where semilocal
solitons satisfy the same equation as \eqref{toda} 
\cite{Kim:1992uw,Kim:1993mh}.

In the present context, we proceed as follows. 
Using \eqref{wz}, we eliminate $u_2(z)$,
\begin{equation} \label{sPopov4}
\partial_z \partial_{\bar z} u_1 = -2(1+|w(z)|^2) e^{u_1},
\end{equation}
where $w(z)$ is an arbitrary rational function of $z$.
If $w(z)=c$ is a constant, which means $\phi_2$ is proportional to $\phi_1$,
\eqref{sPopov4} reduces to the Popov equation of a single scalar. That is,
\eqref{toda} becomes the Liouville equation
of which the exact solutions are well-known \cite{Witten:1976ck}. 
They are given by \cite{Manton:2012fv}
\begin{equation} \label{expu}
|\phi_1|^2 = \frac{(1+|z|^2)^2 |R'(z)|^2}{(1+|c|^2)(1+|R(z)|^2)^2},
\end{equation}
where $R(z)$ is a ratioinal function of $z$. With an appropriate local
gauge choice, $\phi_i$ and $a_{\bar z}$ are then 
\begin{align} \label{sol1}
\phi_1 &= \frac{(1+|z|^2) R'(z)}{\sqrt{1+|c|^2}(1+|R(z)|^2)}, \notag \\
\phi_2 &= c \phi_1, \notag \\
a_{\bar z} 
 &= i \left[ \frac{R(z) \overline{R'(z)}}{1+|R(z)|^2} - \frac{z}{1+|z|^2} 
              \right].
\end{align}
If $R(z)$ is a ratio of polynomials of degree $n$, the vortex number
is $N=2n-2$ which is even. Note that the 
vortex points of $\phi_1$ and $\phi_2$ are the same.

For later use, we express these Liouville solutions in a different form.
Let us write $R(z)$ as a ratio of two polynomials $P(z)$ and $Q(z)$ which
are generically of order $n$ and have no common zeros,
\begin{equation} \label{Rz}
R(z) = \frac{Q(z)}{P(z)}.
\end{equation}
Then \eqref{expu} becomes
\begin{equation} \label{sol12}
|\phi_1|^2 = \frac{(1+|z|^2)^2 |P(z)Q'(z)-Q(z)P'(z)|^2}%
                  {(1+|c|^2)(|P(z)|^2 + |Q(z)|^2)^2},
\end{equation}
and a natural gauge choice would be
\begin{align} \label{gauge}
\phi_1 &= \frac{(1+|z|^2) (P(z)Q'(z)-Q'(z)P(z))}%
               {\sqrt{1+|c|^2}(|P(z)|^2 + |Q(z)|^2)}, \notag \\
\phi_2 &= c \phi_1, \notag \\
a_{\bar z} &= i \left[ \frac{P(z)\overline{P'(z)} + Q(z)\overline{Q'(z)}}%
                         {|P(z)|^2 + |Q(z)|^2} - \frac{z}{1+|z|^2} \right].
\end{align}
This form may be directly obtained from \eqref{sol1} by multiplying
$P^2/|P|^2$ to $\phi_i$ and adding $i\partial_{\bar z} \ln{\bar P}$ 
to $a_{\bar z}$. Of course this is a gauge transformation which removes
singular phases at the vortex points at finite $z$.

For an arbitrary $w(z)$ which is not constant, it turns out to be useful
to introduce
\begin{equation} \label{u1v}
	e^{u_1} = \frac{|w'|^2}{(1+|w|^2)^3}e^v 
\end{equation}
Then \eqref{sPopov4} becomes an equation for $v$,
\begin{equation}
\partial_z \partial_{\bar z} v 
 = \frac{2|w'|^2}{(1+|w|^2)^2} \left( \frac32 - e^v \right).
\end{equation}
Changing the differentiation variable from $z$ to $w$, we get
\begin{equation} \label{Popov32}
\partial_w \partial_{\bar w} v
 = \frac2{(1+|w|^2)^2}\left( \frac32 - e^v \right).
\end{equation}
Note that this equation is identical to the Popov equation \eqref{Popov3}
with the ratio of radii $C=\sqrt{3/2}$. In other words,
the semilocal Popov equation 
with equal radii reduces to the Popov equation 
with $C=\sqrt{3/2}$.

As shown in \cite{Popov:2012av}, \eqref{Popov32} is not integrable and
general solutions are not known. Nevertheless there is one known exact 
solution, namely $e^v = 3/2$. Though it is a trivial solution by itself,
it provides nontrivial solutions to the semilocal Popov equation
under consideration. From \eqref{uphi} and \eqref{u1v}, we see that
\begin{equation} \label{solphi12}
|\phi_1|^2 = \frac32 \frac{(1+|z|^2)^2 |w'|^2}{(1 + |w|^2)^3}
\end{equation}
solves the semilocal Popov equation \eqref{sPopov3}.
Given $|\phi_1|^2$, we can construct other fields. 
At this time $w(z)$ plays the role of $R(z)$ of \eqref{expu} and 
is written as a ratio of two polynomials $P(z)$ and $Q(z)$ with no common zeros,
\begin{equation} \label{wpq}
w(z) = \frac{Q(z)}{P(z)}.
\end{equation}
Then we find
\begin{equation} \label{solphi2}
\begin{pmatrix} |\phi_1|^2 \\ |\phi_2|^2 \end{pmatrix}
 = \frac32 \frac{(1+|z|^2)^2 |P(z)Q'(z)-Q(z)P'(z)|^2}%
                             {(|P(z)|^2 + |Q(z)|^2)^3}
\begin{pmatrix} |P(z)|^2 \\ |Q(z)|^2 \end{pmatrix}.
\end{equation}
With a local gauge choice as in \eqref{gauge}, the solutions are then given by
\begin{align} \label{finalsol}
\begin{pmatrix}
	\phi_1 \\ \phi_2
\end{pmatrix}
&=
\sqrt{\frac32} \frac{(1+|z|^2) (P(z)Q'(z)-Q(z)P'(z))}%
	                        {(|P(z)|^2 + |Q(z)|^2)^{3/2}}
\begin{pmatrix}
	P(z) \\ Q(z)
\end{pmatrix}, \notag \\
a_{\bar z} &= i \left[ 
  \frac32 \frac{P(z)\overline{P'(z)} + Q(z)\overline{Q'(z)}}%
	       {|P(z)|^2 + |Q(z)|^2} - \frac{z}{1+|z|^2} \right].
\end{align}
Note that $\phi_1$ vanishes at the zeros of $P$ and $PQ'-QP'$, while
$\phi_2$ vanishes at the zeros of $Q$ and $PQ'-QP'$. Thus they share
only part of the vortex points and this is genuinely different family
of solutions from the Liouville solutions \eqref{gauge}. 

Let us count the vortex number of the solution \eqref{finalsol} 
for a generic rational function $w$ of degree $n$ with $P$ and 
$Q$ being polynomials of order $n$. Since $PQ'-QP'$ is a polynomial
of order $2n-2$, $\phi_1$ and $\phi_2$ have $3n-2$ zeros respectively.
In addition, we have to consider the behavior at $z=\infty$ but it is
easy to see that the scalar fields remain finite at $z=\infty$. 
Thus the vortex number is $N=3n-2$. 
We can also confirm the result by investigating the singularities
of the solution. With the local gauge choice in \eqref{finalsol},
$\phi_i$'s are regular everywhere except at $z=\infty$. In particular,
around a zero $z_0$ of $\phi_i$, $\phi_i$ behave as $\phi_i \sim c(z-z_0)$. 
The expression \eqref{finalsol} however is singular at $z=\infty$ since
\begin{equation}
\phi_i \sim c \frac{z^{3n-2}}{|z|^{3n-2}},
\end{equation}
This can be removed by a gauge transformation of winding number $3n-2$
defined on an annulus on $S^2$ enclosing $z=\infty$. The winding number
is then the first Chern number $N$, which is the same as the vortex
number. If the order of $P$ or $Q$ is not $n$ but less than $n$, there 
would be vortices sitting at $z=\infty$ but it is easy to see that the 
vortex number is still $N=3n-2$. Note that solutions with odd vortex 
numbers are possible in contrast to the Liouville solutions \eqref{gauge}
which have even vortex numbers $N=2n-2$. Moreover the solutions with
unit vorticity belong to this family.

As an example of the solution, choose $P(z) = c^n$ and $Q(z) = z^n$ with
$c>0$ for simplicity. Then
\begin{equation} \label{nsolution}
\begin{pmatrix}
	\phi_1 \\ \phi_2
\end{pmatrix}
= \sqrt{\frac32} \frac{nc^n (1 + |z|^2) z^{n-1}}%
                       {(c^{2n} + |z|^{2n})^{3/2}}
   \begin{pmatrix} c^n \\ z^n \end{pmatrix}, 
\end{equation}
This is a circular symmetric solution with vortices at $z=0$
with multiplicities $n-1$ and $2n-1$ for $\phi_1$ and $\phi_2$, respectively.
The constant $c$ may be considered as a parameter representing the size
of the vortices. In terms of the coordinate $\xi = 1/z$, this becomes
\begin{equation} \label{nsolution2}
\begin{pmatrix}
	\phi_1 \\ \phi_2
\end{pmatrix}
= \sqrt{\frac32} 
   \left(\frac{|\xi|}{\xi}\right)^{3n-2}
   \frac{nc^{-n} (1 + |\xi|^2) \xi^{n-1}}%
                       {(c^{-2n} + |\xi|^{2n} )^{3/2}}
   \begin{pmatrix} \xi^n \\ c^{-n} \end{pmatrix},
\end{equation}
where $(|\xi|/\xi)^{3n-2}$ is the phase factor mentioned above and
should be removed by a gauge transformation. The exponent $3n-2$ is
the first Chern number which should be the total vortex number. 
Indeed there are vortices at $z=\infty$ with
multiplicities $2n-1$ and $n-1$ for $\phi_1$ and $\phi_2$, respectively,
as seen in \eqref{nsolution2}.
The size of the vortices at $z=\infty$ is $1/c$. For $c=1$, the solution
has the reflection symmetry in the equator $|z|=1$ if $\phi_1$ and
$\phi_2$ are exchanged.

The solution with unit vorticity $N=1$ and $c=1$ is worth mentioning
separately. With $P(z) = 1$ and $Q(z)=z$, the conformal factor $1+|z|^2$
in the numerator is cancelled by the denominator in \eqref{nsolution} 
and we get
\begin{equation} \label{unitvortex}
\begin{pmatrix}
	\phi_1 \\ \phi_2
\end{pmatrix}
= \sqrt{\frac32} \frac1{\sqrt{1 + |z|^2}}
   \begin{pmatrix} 1 \\ z \end{pmatrix}. 
\end{equation}
This is precisely the well-known $CP^1$ lump configuration with 
unit size \cite{D'Adda:1978uc}. The scalars satisfy
\begin{equation} \label{s3}
|\phi_1|^2 + |\phi_2|^2  = \frac32,
\end{equation}
which defines $S^3$ fibered as a circle bundle over $CP^1$.
Note that the radius of $S^3$ is not one but $\sqrt{3/2}$.
The origin of this radius may be traced back to the connection to the
Popov equation with ratio of radii $C=\sqrt{3/2}$ as seen in \eqref{Popov32}.
From \eqref{finalsol} the gauge potential of the solution is calculated as
\begin{equation}
a_{\bar z} = \frac{i}2 \frac{z}{1 + |z|^2},
\end{equation}
and the corresponding field strength is
\begin{equation}
F_{z \bar z} = \frac{i}{(1 + |z|^2)^2},
\end{equation}
which can also be directly obtained from \eqref{sPopov2}. The
magnetic charge is uniform on $S^2$ and is that of the Dirac monopole with 
unit magnetic charge. Considering the form of \eqref{sPopov2}, it is clear
that the relation \eqref{s3} is crucial to obtain the magnetic field with
unit magnetic charge.

In conclusion, we have considered the semilocal version of Popov's vortex
equations on $S^2$. Though they are not integrable, we were
able to construct two families of exact solutions both of which involve 
arbitrary rational functions on $S^2$. One family of solutions is just
a trivial embedding of the solutions of the Popov equations and
they have even vortex numbers $N=2n-2$, where $n$ is the degree of the
rational function. The other family of solutions have
vortex number $N=3n-2$ and they correspond to the constant solution with
Popov equation with the ratio of radii $C=\sqrt{3/2}$. The $N=1$
solution with the reflection symmetry is given by the $CP^1$ lump
configuration with unit size. Obviously there should be other
solutions. This is evident if we recall that the solutions found here
are just obtained from the constant solution of \eqref{Popov32}.
A related issue is that we have even vortex numbers for even $n$. 
Thus for the vortex numbers $N=6k-2$ ($k\in \mathbb{Z}$), 
we have two distinct families of solutions: Liouville-type solutions
are generated by rational functions of degree $3k$ while the other
solutions, by those of degree $2k$. A natural question is then
whether the solutions are smoothly connected to each other in the
solution space through solutions not found here. To answer these
questions it may be helpful to do the zero-mode analysis similar
to that in \cite{Weinberg:1979er}.
In this paper, we have shown that the semilocal Popov equations are obtained
as Bogomolny equations in the nonrelativistic Chern-Simons matter systems 
on $S^2$ in the presence of a constant external magnetic field. 
The Popov equations however are originally discussed as a reduction
of Yang-Mills instanton equations with noncompact gauge group. It would
be interesting to find such an interpretation for the semilocal equations
for example as an embedding in nonabelian Popov equations.

\section*{Acknowledgements}
This work was supported by the Mid-career Researcher Program
grant No.\ 2012-045385/2013-056327 and the WCU grant No.\ R32-10130
through NRF of Korea funded by the Korean government (MEST), 
and the Research fund No.\ 1-2008-2935-001-2 by Ewha Womans University.

\end{document}